\documentclass[12pt,preprint]{aastex}
\usepackage{graphicx}
\usepackage{emulateapj5}
\usepackage{apjfonts}
\usepackage{epsfig}
\usepackage{natbib}
\def\gtrsim{\mathrel{\hbox{\rlap{\hbox{\lower4pt\hbox{$\sim$}}}\hbox{\raise2pt\hbox{$>$}}}}}

\newcommand{\hst}{\emph{HST}}

\newcommand{\kms}{km~s\ensuremath{^{-1}}}

\newcommand{\mbh}{\ensuremath{M_\mathrm{BH}}}

\newcommand{\msigma}{\ensuremath{M_{\mathrm{BH}}-\sigmastar}}
\newcommand{\msun}{\ensuremath{M_{\odot}}}

\newcommand{\sers}{S{\'e}rsic}

\newcommand{\sigmastar}{\ensuremath{\sigma_{\ast}}}

\def\lax{{$\mathrel{\hbox{\rlap{\hbox{\lower4pt\hbox{$\sim$}}}\hbox{$<$}}}$}}
\def\gax{{$\mathrel{\hbox{\rlap{\hbox{\lower4pt\hbox{$\sim$}}}\hbox{$>$}}}$}}

\begin{document}

\title{Megamaser Disks Reveal a Broad Distribution of Black Hole Mass in Spiral Galaxies}

\author{J.~E.~Greene\altaffilmark{1}, A.~Seth\altaffilmark{2}, 
M. Kim\altaffilmark{3}, R. L{\"a}sker, A. Goulding\altaffilmark{1}, 
F. Gao\altaffilmark{6}, J.~A.~Braatz\altaffilmark{4},
C.~Henkel\altaffilmark{5}, J. Condon\altaffilmark{4}, K.~Y.~Lo\altaffilmark{4},
W.~Zhao}

\altaffiltext{1}{Department of Astrophysics, Princeton University}
\altaffiltext{2}{University of Utah, Salt Lake City, UT 84112, USA}
\altaffiltext{3}{Korea Astronomy and Space Science Institute, Daejeon 305-348, Korea; University of Science and Technology, Daejeon 305-350, Korea}
\altaffiltext{4}{National Radio Astronomy Observatory, 520 Edgemont Road, Charlottesville, VA 22903, USA}
\altaffiltext{5}{Max-Planck-Institut f{\"u}r Radioastronomie, Auf dem H{\"u}gel 69, D-53121 Bonn, Germany; Astronomy Department, King Abdulaziz University, P.O. Box 80203, Jeddah 21589, Saudi Arabia}
\altaffiltext{6}{Key Laboratory for Research in Galaxies and Cosmology, Shanghai Astronomical Observatory, Chinese Academy of Science, Shanghai 200030, China; National Radio Astronomy Observatory, 520 Edgemont Road, Charlottesville, VA 22903, USA; Graduate School of the Chinese Academy of Sciences, Beijing 100039, China}

\begin{abstract}
We use new precision measurements of black hole masses from water
megamaser disks to investigate scaling relations between macroscopic
galaxy properties and supermassive black hole (BH) mass. The
megamaser-derived BH masses span $10^6-10^8$~\msun, while all the
galaxy properties that we examine (including total stellar mass,
central mass density, and central velocity dispersion) lie within a
narrower range. Thus, no galaxy property correlates tightly with
\mbh\ in $\sim L^*$ spiral galaxies as traced by megamaser disks.  Of
them all, stellar velocity dispersion provides the tightest relation,
but at fixed \sigmastar\ the mean megamaser \mbh\ are offset by $-0.6
\pm 0.1$ dex relative to early-type galaxies. Spiral galaxies with
non-maser dynamical BH masses do not appear to show this offset. At
low mass, we do not yet know the full distribution of BH mass at fixed
galaxy property; the non-maser dynamical measurements may miss the
low-mass end of the BH distribution due to inability to resolve their
spheres of influence and/or megamasers may preferentially occur in
lower-mass BHs.
\end{abstract}

\section{Introduction}

For more than a decade, we have studied the relationship between the
stellar velocity dispersion \sigmastar\ of a galaxy bulge and the
central supermassive black hole (BH) mass
\mbh\ \citep[e.g.,][]{tremaineetal2002}.  As more
dynamical \mbh\ are measured, particularly at the highest
\mbh\ \citep[e.g.,][]{mcconnellma2013} and in spiral galaxies
\citep{greeneetal2010}, it has become clear that the
relationship between \mbh\ and \sigmastar\ has more scatter and more
structure than was originally thought \citep{kormendyho2013}.

BH mass does not correlate strongly with total galaxy mass
\citep[e.g.,][]{reinesvolonteri2015}, nor does it appear to correlate with
circular velocity \citep[and thus by inference dark matter halo
  mass;][]{sunetal2013}. It is possible that \mbh\ is built up along
with the bulge mass. However, in late-type galaxies where there is
clear evidence for ongoing bulge growth, \mbh\ appear to be
systematically lower than expected based on BH-bulge relations in
early-type galaxies \citep[e.g.,][]{laeskeretal2016}. Here,
we use megamaser disks that reside well within the BH sphere of
influence to probe the distribution of \mbh\ in spiral galaxies.

We assume an H$_0$ of 70 km~s$^{-1}$~Mpc$^{-1}$
\citep{dunkleyetal2009}, consistent with all of the H$_0$
measurements from megamaser disk galaxies measured to date
\citep[e.g.,][]{kuoetal2015}.

\section{Megamaser Disk Galaxies, Sample, and Observations}
\label{sec:Sample}

Stellar and non-maser gas dynamical methods yield the majority of
dynamical \mbh, but they are limiting both in spatial resolution 
and in ability to cleanly measure \mbh\ in all galaxy types. 
Due to the need to resolve the gravitational
sphere of influence, these methods currently cannot reach \mbh\ 
$< 10^7$~\msun\ at the distance of Virgo.
Dust, ongoing star formation, and non-axisymmetric potentials such as
bars also complicate efforts to get \mbh\ in late-type galaxies.
Megamaser disks circumvent these problems and provide the most
accurate and precise extragalactic \mbh\ measurements.

Circumnuclear H$_2$O megamasers often trace Keplerian rotation around the
supermassive BH at radii of $\sim 0.2-1$ pc
\citep[e.g.,][]{miyoshietal1995}. Sensitive surveys, enabled
primarily by the Green Bank Telescope (GBT), and micro-arcsecond
accurate mapping of the disks with the Very Long Baseline Array 
(VLBA), the Jansky Very Large Array (JVLA), 
the GBT, and/or the 100-m Effelsberg telescope have yielded 
more than 150 megamaser galaxies, of which at least 34
harbor megamaser disks \citep[e.g.,][]{pesceetal2015}.  The host galaxies
are $\sim L^*$ spirals with Hubble types S0-Sbc, while the
derived \mbh\ have values ranging from $10^6-10^8$~\msun\
 \citep[e.g.,][]{greeneetal2010,kuoetal2011}. Here we add
\sigmastar\ measurements for seven galaxies, whose \mbh\ are
presented in \citet{gaoetal2016}, Gao et al.\ in preparation, 
and Zhao et al.\ in preparation (Table 1).

\subsection{Comparison Galaxy Sample}

We consider only galaxies with dynamical \mbh\ measurements. We start
with the recent compilation from \citet[][97 galaxies]{sagliaetal2016}
plus a new megamaser disk mass for IC2560 (Wagner et al. in
preparation).  Of these 98, 13 are megamaser disks.  We add non-maser
dynamical masses for NGC 4395 \citep{denbroketal2015}, and the
interesting outliers NGC 1271 \citep{walshetal2015}, NGC 1277
\citep{walshetal2016}, and NGC 1600 \citep{thomasetal2016} for a total
of 102 literature sources and a final sample of 109 galaxies including
our 7 new megamaser disks (Table 1). There are a total of 20 megamaser
disk galaxies and 17 late-type (non-S0) non-maser spiral
galaxies. When we discuss stellar masses ($M_{*}$), we refer to the 67
galaxies that have imaging from the Sloan Digital Sky Survey
\citep[SDSS;][]{yorketal2000}.

\subsection{Spectroscopic Observations}

\vbox{ 
\vskip -2mm
\hskip +1mm
\includegraphics[width=0.4\textwidth]{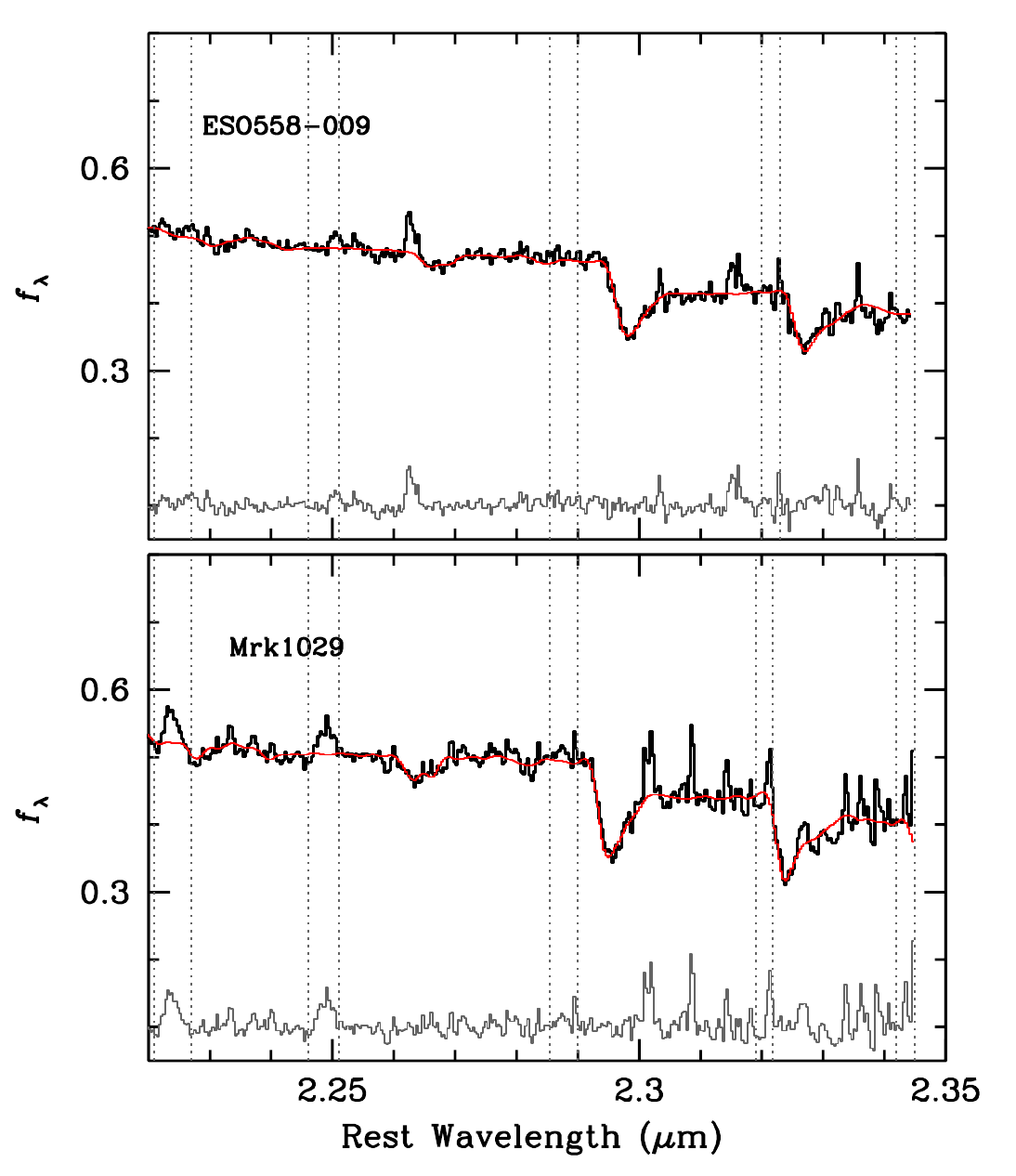}
}
\vskip -0mm
\figcaption[]{
Fits to the observations of ESO 558$-$009 and Mrk 1029 from Triplespec 
(\S 2.2). 
The data are shown in black 
(in $f_{\lambda}$ units of erg~s$^{-1}$~cm$^{-2}$~\AA$^{-1}$ arbitrarily scaled), 
the best-fit pPXF model in red, and residuals in thin black at the bottom.
Masked regions corresponding to emission lines are indicated with vertical 
dashed lines. Unmasked spikes are the result of imperfect sky subtraction.
\label{fig:twospec}}
\vskip 5mm

The \sigmastar\ presented here are measured from three data sets. Two
galaxies (NGC 1320, NGC 5495) were observed with the B\&C spectrograph
on the Dupont telescope at Las Campanas Observatory using a
2\arcsec\ slit \citep{greeneetal2010}. These spectra have an
instrumental resolution of $\sigma_r \approx 120$~\kms\ and a
wavelength range of $3400-6000$\AA. NGC 1320 is only marginally
spectroscopically resolved, while NGC 5495 is well-resolved, but we
are able to recover reliable dispersions at or slightly below the
instrumental resolution \citep{greeneetal2010}. We extracted spectra
within a 2\arcsec\ aperture to perform our measurements.  The
reductions are described in detail in \citet{greeneetal2010}.
Standard flat-field, bias, wavelength calibration, flux, and telluric
correction were performed within IRAF.  Each galaxy was observed
  for 1.3 hr, yielding signal-to-noise ratios of $\sim 100$
  pixel$^{-1}$ (see example fits to similar objects in Greene et
  al.\ 2010).

Two galaxies (Mrk1029, ESO558$-$G009) have
\sigmastar\ measurements from the cross-dispersed near-infrared
spectrograph Triplespec \citep{wilsonetal2004} on the 3.5m telescope
at Apache Point. Triplespec has a wavelength range of $0.9-2.4
\micron$ with a spectral resolution of $\sigma_r \approx 37$~\kms, far 
higher than the internal \sigmastar\ of the galaxies.  We
observed with the $1\farcs1$ slit oriented along
the major axis of each galaxy; again we extracted data within a
  2\arcsec\ aperture.  We nodded off the galaxy by 1-3\arcmin\ to
measure the sky, and at least once an hour we observed an A0V star
($10 < H < 6$ mag) at similar airmass to serve as a spectrophotometric
standard and to facilitate telluric correction.
Integration times ranged from 0.5-1.5 hr, yielding signal-to-noise 
ratios of $\sim 50$ pixel$^{-1}$ (Fig.\ \ref{fig:twospec}).

\vbox{ 
\vskip 2mm
\hskip -2mm
\includegraphics[width=0.45\textwidth]{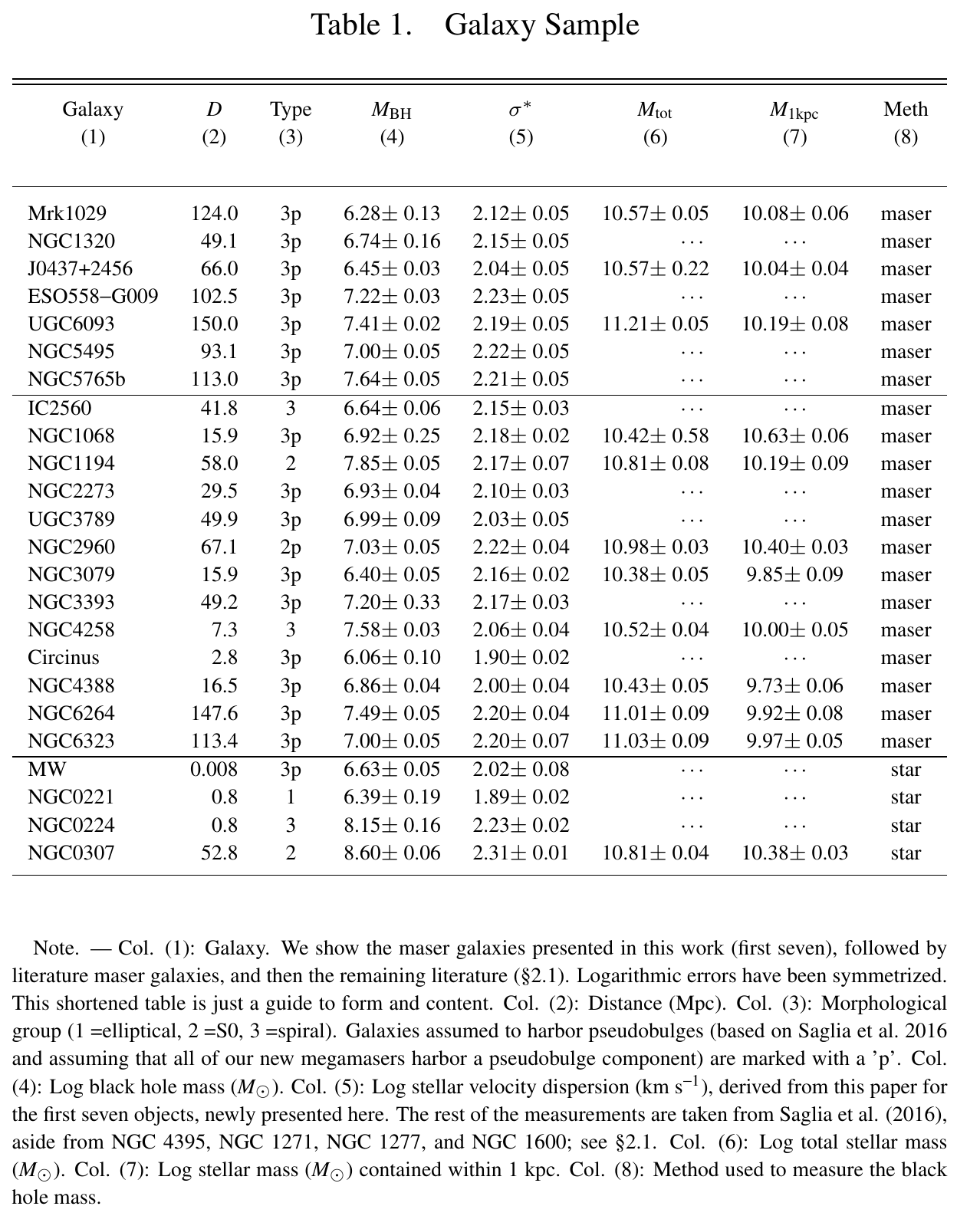}
}
\vskip 5mm

We performed standard data reduction procedures using custom software
written for Triplespec based on the Spextool package
\citep{vaccaetal2003,cushingetal2004}, which performed flat-fielding,
wavelength calibration, bias, dark, and atmospheric air-glow
subtraction. Nonlinearity corrections were applied, and the spectra are
traced and optimally extracted \citep{horne1986}.  The spectra were
combined using a robust error clipping at each pixel. Flux calibration
utilized observations of A0 telluric standards \citep{vaccaetal2003}.

Finally, three galaxies (J0437+2456, NGC 5765b, UGC 6093) have spectra
from the SDSS. They have a spectral resolution of
$\sigma_r\approx 65$~\kms\ (resolving our galaxies well) and cover a
range of $3800-9200$\AA.  The fiber has a 3\arcsec\ diameter,
roughly matched to the extraction apertures used above.

\begin{figure*}
\vbox{ 
\vskip -2mm
\hskip +10mm
\includegraphics[width=0.8\textwidth]{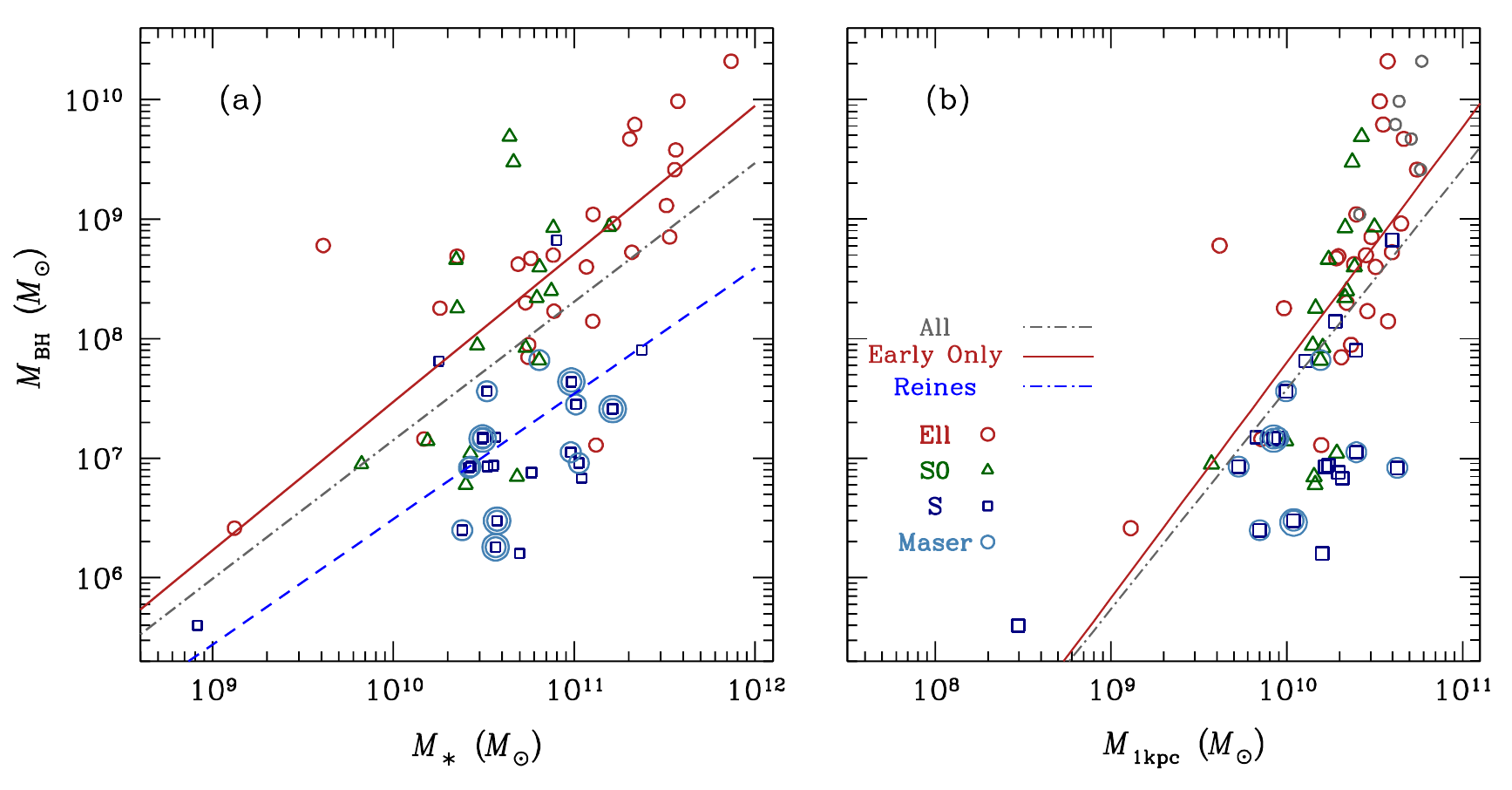}
}
\vskip -0mm
\figcaption[]{
({\bf a}): Relationship between \mbh\ and $M_{\ast}$. 
We show elliptical (red circles), 
S0 (green triangles), spiral (blue squares), and megamaser disk 
(blue circles) galaxies. Double circles indicate our new measurements. We 
show fits to the full sample (grey dot-dashed), the early-types (red 
solid) and the \citet{reinesvolonteri2015} fit (blue dashed).
({\bf b}): The relationship between \mbh\ and mass enclosed 
within 1 kpc. Symbols as at left. In grey we 
show the galaxy mass ``corrected'' for core-scouring 
\citep[e.g.,][]{ruslietal2013core}. 
Galaxies with radii $<$2'' are excluded.
\label{fig:bhcent}}
\end{figure*}

\section{Analysis}
\label{sec:analysis}

\subsection{Stellar Velocity Dispersions}

We measure \sigmastar\ using direct pixel fitting implemented by the
publicly available pPXF code \citep{cappellariemsellem2004}. We
measure only the first two moments of the line-of-sight velocity
distribution ($V$ and \sigmastar). To account for differences in flux
calibration and reddening, we also include a fourth-order
multiplicative polynomial.

For the two optical spectrographs, we use spectral templates from
\citet{valdesetal2004}, which have a higher spectral resolution than
the galaxy observations ($\sigma_r \approx 24$~\kms). The
code solves for the optimal set of template weights over the
spectral range of $4000-5400$\AA. We experiment with three
other narrower spectral windows, and find agreement within $\sim 20\%$
for \sigmastar\ between the different regions.

Our near-infrared spectral template stars are taken from
\citet{wallacehinkle1996} and have a spectral resolution of 
$\sim 33$~\kms\ in the $K$-band.  The NIR spectra cover $YJHK$. 
We derive the most stable \sigmastar\ measurements
from the CO bandheads in the $K$-band
(2.2-2.35 $\micron$). The $H-$band data proved hard to fit due to
spectral differences between the galaxies and our template stars.

\subsection{Uncertainties in \sigmastar}

The errors on \sigmastar\ are measured via bootstrapping; the best-fit
models are combined with Gaussian-random noise of the proper amplitude
to simulate the data and these are refit to generate a distribution in
$V$ and \sigmastar\ (Table 1). We also compare our observations with
the \sigmastar\ measurements published by the Hobby Eberly Telescope
Massive Galaxy Survey \citep[HETGMS;][]{vandenboschetal2015}.  Taking
the eight megamaser galaxies in common between HETMGS and the
\sigmastar\ published in Greene et al.\ (2010) or presented here, we
find $(\langle \sigma_{\rm HETMGS} - \sigma_{\rm This
  work})/\sigma_{\rm HETMGS} \rangle =-0.03 \pm 0.13$.  We set 13\% as
a floor on the measured uncertainty.
 
Aperture corrections are a significant source of systematic
uncertainty.  In ellipticals, \sigmastar\ is typically measured within
a fixed fraction of $R_e$, which is challenging to measure in spiral
bulges. Furthermore, \sigmastar\ profiles are more complex and
$V/\sigma$ is typically higher in spiral bulges, even when the
large-scale disk is excluded, as here
\citep[e.g.,][]{falconbarrosoetal2006,greeneetal2014}.  We do not 
correct \sigmastar\ for rotation, but we
estimate the magnitude of such a correction as follows. Typical spirals have
$V/\sigmastar \approx 0.3$ within
1-2\arcsec\ \citep{falconbarrosoetal2006}. Removing the rotation in
quadrature moves the spirals toward the
\msigma\ relation of ellipticals \citep[e.g.,][]{bennertetal2015}.
On the other hand, elliptical galaxies also
typically rotate \citep[e.g.,][]{emsellemetal2007} so there is no
strong scientific justification to remove this component of dynamical 
support only in the spiral galaxies.

\subsection{Stellar Masses}

In addition to \sigmastar, we calculate stellar masses ($M_{\ast}$)
for the 67 galaxies that lie within the SDSS footprint using linear
relationships between color and $M/L$ from \citet{belletal2003}. To
estimate an allowed range of $M_{\ast}$, for each galaxy we calculate
12 possible stellar masses using the four SDSS Petrosian colors to
derive $M/L$, combined with the robustly measured $g$, $r$, and
$i-$band luminosities.  We average these estimates and take the
standard deviation to estimate an error -- this error is larger in
systems with a wide mix of stellar populations and/or dust, where 
blue and red colors provide quite different $M/L$ estimates.

We calculate $M_{\ast}$ for the entire galaxy and within $r < 1$~kpc
($M_{\rm 1 kpc}$).  The mass within a fixed physical aperture is a
crude but non-parametric estimate of the central stellar density, and
might correlate better with \mbh\ than the total $M_{\ast}$.  It is
also a simple measurement relative to $M_{\rm bulge}$,
\sigmastar, or even $M_{\rm \ast}$.  We measure $M_{\rm 1 kpc}$ from
the SDSS data using the luminosity within a central circular aperture.
We exclude galaxies where the aperture is $<$2'', which are limited by
the SDSS seeing.

\subsection{Mass Uncertainties}

Photometric stellar masses contain many uncertainties, including
aperture effects, projection effects, and the contribution of nebular
emission from an active nucleus (for the megamaser disk galaxies).
The latter effect we mitigate by taking an average over multiple
filters.  We find only small differences when we test elliptical
apertures for a subset of the galaxies ($\sim 0.2$ mag), and removing
edge-on galaxies does not change the result. We also explore whether
our results are biased by using projected quantities. For the
high-mass, high-\sers\ index galaxies, we analytically deproject the
central mass \citep{bezansonetal2009} using the single-\sers\ fits
from the NSA catalog.  We reproduce our aperture measurements within
0.1 dex, with no systematic dependence on mass or $n$. Deprojection 
uncertainties may play a larger role for the pseudobulges, 
particularly for nearly edge-on inclinations. We explore the importance of 
the inclination using ellipse fitting to our \hst\ data for the 
maser galaxies \citep{greeneetal2014,laeskeretal2016} and find that 
the differences between circular and elliptical apertures are never larger
than 0.2 mag, translating into $<0.1$ dex in stellar mass.

Hardest to quantify are the errors incurred from our mass estimation
techniques, but overall it has been shown that single-color
conversions (\S 3.3) return an unbiased stellar mass
\citep[][]{roedigercourteau2015}.

\section{Exploring Black Hole-Galaxy Correlations}

In Figure \ref{fig:bhcent}a, we show the relationship between
\mbh\ and $M_{\ast}$.  We fit all relations in this
paper using the Bayesian line-fitting code of \citet{kelly2007},
taking into account uncertainties in both \mbh\ and the galaxy
properties. We fit a log-linear model with log ($\mbh/\msun) =
\alpha\,+\,\beta \, \rm{log} \, (X/X_0) + \epsilon$, with 
$X$ being $M_{\ast}$, $M_{\rm 1 kpc}$, or \sigmastar\ and $X_0$ 
near the center of the distribution (Table 2).

We find $\sim 1$ dex scatter in the relation between $M_{\ast}$ and
\mbh. At a stellar mass of $\sim 10^{11}$~\msun, \mbh\ spans a range
from $10^6$ to $10^9$~\msun, and spiral galaxies tend to
have lower \mbh\ than elliptical galaxies at
similar $M_{\ast}$. We have long known that BHs correlate better with
galaxy bulges than with $M_{\ast}$, disk mass
\citep[e.g.,][]{gebhardtetal2003}, or asymptotic circular velocity 
\citep[e.g.,][]{sunetal2013}.

Original interpretations of the BH-bulge scaling relations posited
that energy released during active phases of BH growth acted on
galaxy-wide gas reservoirs, thus impacting the future growth of the
galaxy \citep[e.g.,][]{springeletal2005}.  The BHs in the megamaser
disk galaxies are too large to be primordial seeds
\citep[e.g.,][]{volonteri2010}, so they have certainly undergone some
accretion episodes, but the lack of correlation with $M_{\ast}$
strongly suggests that BH growth does not remove gas from galactic
disks (i.e., does not act beyond bulge scales).  After removing the
disk component, we still see considerable scatter between \mbh\ and
bulge properties in late-type spiral galaxies, even when we attempt to
isolate the ``classical'' bulge component from the nuclear
disks, rings, and bars that characterize ``pseudobulges''
\citep{laeskeretal2016,sagliaetal2016}.

\subsection{A Local Relationship with Central Density}

Recent work has found that central mass density (stellar mass within 1
kpc, $M_{\rm 1 kpc}$) is a strong indicator of the star formation
history of a galaxy \citep[e.g.,][]{barroetal2015}, with high central
stellar density and galaxy quenching going hand in hand.  Exactly what
terminates star formation in dense galaxies remains unknown, but one
possibility is feedback from supermassive BHs, which may grow along
with dense cores.  Thus, we consider a local relationship between
\mbh\ and the mass enclosed within 1 kpc (Figure \ref{fig:bhcent}b).

A local relationship between \mbh\ and stellar mass is cheap to
measure (\S 3.3) and relatively easy to compare with simulations
\citep[see][]{reinesvolonteri2015}.  Such a local relationship might
arise if the gas supply to the galaxy center determines both the
central star formation rate and the overall rate of BH growth, and BH
growth is self-regulated \citep[e.g.,][]{debuhretal2010}, or if the
accretion disk is fed locally by stars rather than gas accretion
\citep{miraldaescudekollmeier2005}.

The \mbh-$M_{\rm 1 kpc}$ relation is shown in Figure
\ref{fig:bhcent}b. We do observe an overall correlation between
$M_{\rm 1 kpc}$ and \mbh. However, the scatter is substantial
(0.8 dex; Table 2), and the dynamic range in $M_{\rm 1 kpc}$ is
limited, causing a very steep scaling between $M_{\rm 1 kpc}$ and
\mbh. At the highest masses, core scouring
\citep[e.g.,][]{ruslietal2013core} tends to artificially decrease the
central mass density. We ``correct'' the most massive galaxies for
core scouring in Fig. \ref{fig:bhcent}b by adding the \mbh\ to the
central mass, but our fits do not change. 

\vbox{ 
\vskip -2mm
\hskip +1mm
\includegraphics[width=0.40\textwidth]{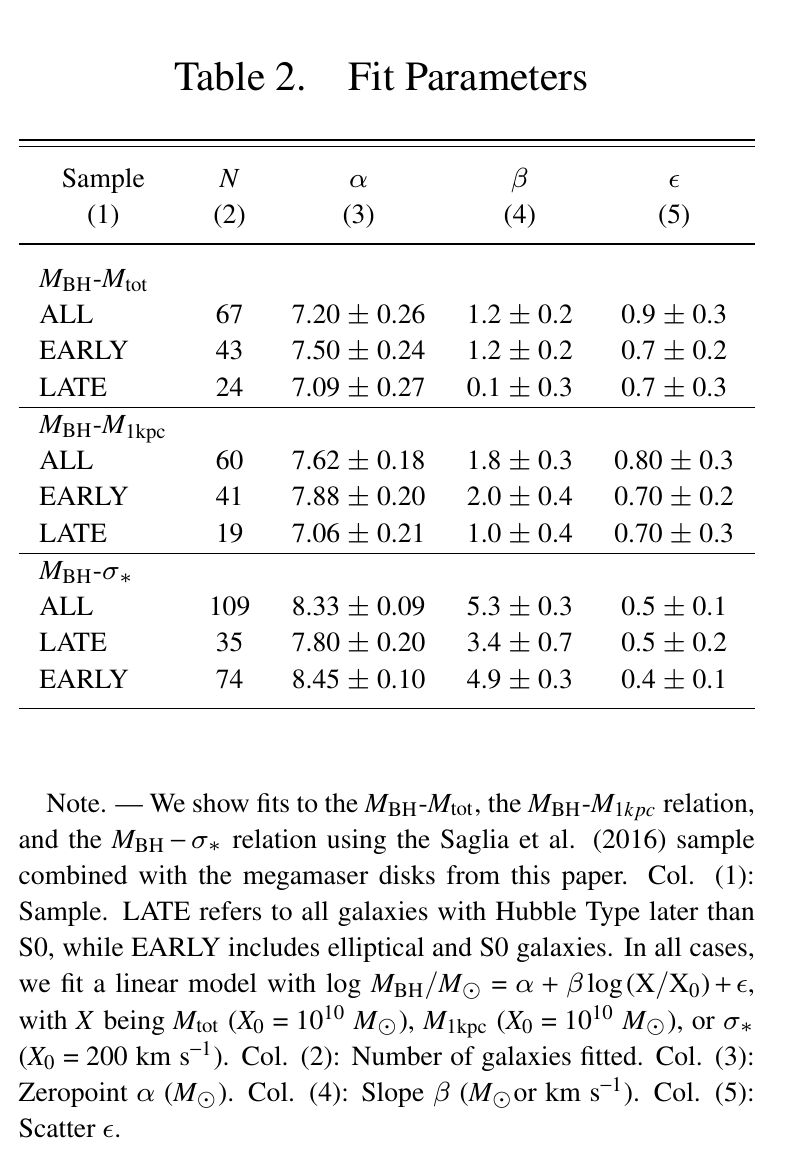}
}
\vskip 5mm

\subsection{Black Hole Mass and Stellar Velocity Dispersion}

Turning to the \mbh-\sigmastar\ relation, our linear fit to the full
sample of galaxies confirms the steep slope found in recent works
($\beta = 5.3 \pm 0.3$) and the high intrinsic scatter ($\epsilon =
0.5 \pm 0.1$).  In \citet{greeneetal2010}, we found that the BH masses
in the megamaser disk galaxies are lower than expected at a fixed
\sigmastar. Here we double the megamaser disk sample and strongly
confirm our previous result (Fig.\ \ref{fig:msigma}). We calculate the
average offset $\Delta M_{\rm BH}$ between \mbh\ and that predicted by
the \msigma\ relation fit to early-type galaxies.  The offset is
calculated as the weighted mean of the difference between the measured
\mbh\ and that calculated from the best-fit relation, with the weights
including both $y$ and $x$ uncertainties combined in quadrature
\citep{laeskeretal2016}.  We find an offset of $-0.39 \pm 0.12$ for
late-type galaxies and an offset of $-0.45 \pm 0.12$ for
pseudobulges\footnote{Taking the identifications from Saglia et
  al.\ 2016 for the literature sources and taking all of our new
  galaxies to harbor some secularly evolving component based on our
  new \hst\ imaging; Pjanka et al. in preparation.}.

We now have marginally sufficient statistics to compare the
distributions of maser and non-maser spirals. There are 20 megamaser
disk galaxies (2 S0, 18 spiral) and 17 late-type (non-S0) spiral
galaxies with \mbh\ measurements from non-maser dynamics.  The maser
and non-maser samples have indistinguishable distributions in
\sigmastar\ according to an Anderson-Darling test
\citep[e.g.,][]{babufeigelson2006} with $p=0.6$ of being drawn from
the same distribution.  Likewise, the distributions of bulge type are
quite similar, with $\sim 75$\% of the stellar/gas dynamical and $\sim
85$\% of the megamaser disk galaxies hosted by pseudobulges (Table 1).

\vbox{ 
\vskip +1mm
\hskip +1mm
\includegraphics[width=0.45\textwidth]{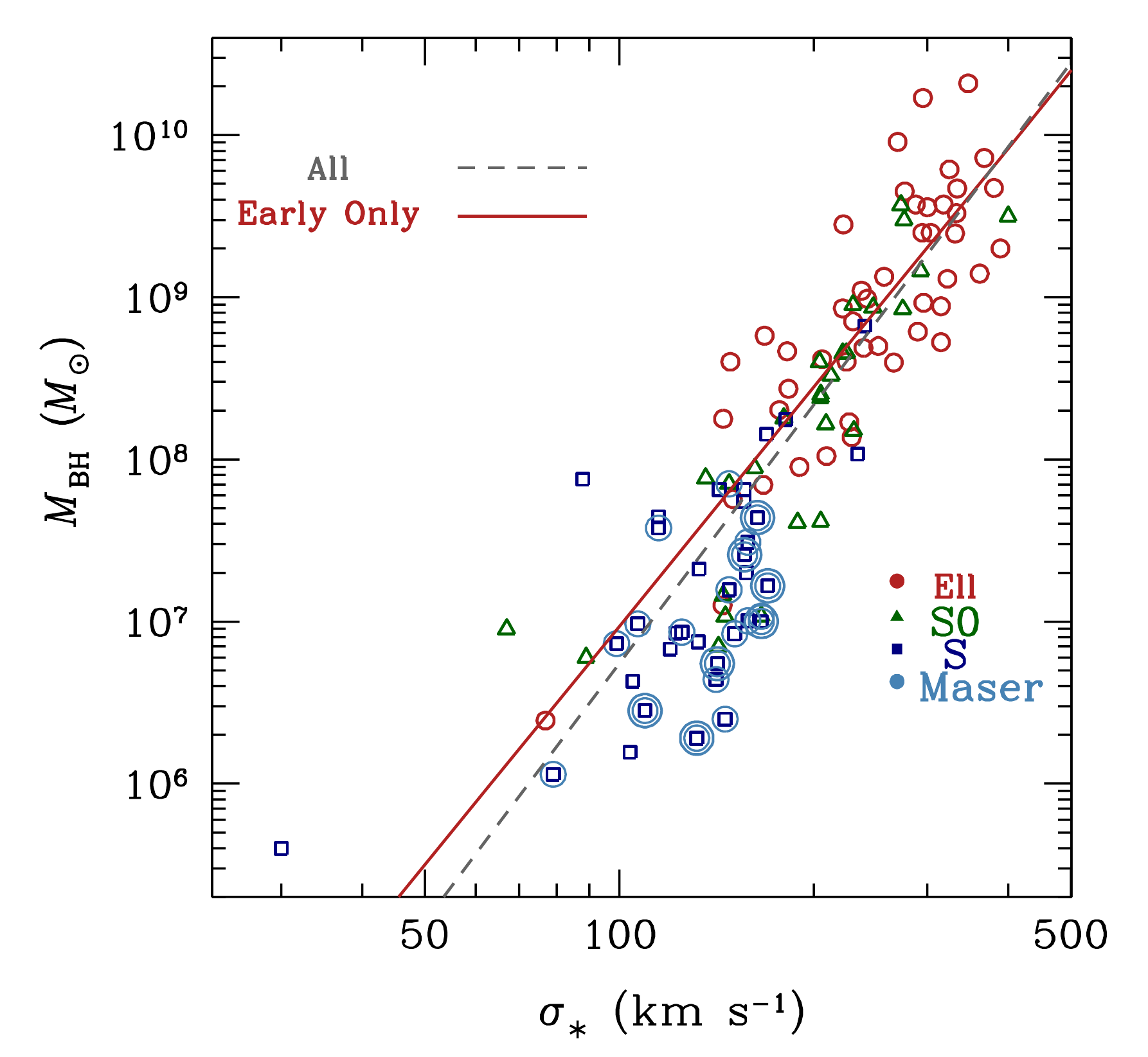}
}
\vskip -2mm
\figcaption[]{
The relationship between \sigmastar\ and \mbh. 
We fit the entire sample (grey dashed line)
and the early-type galaxies alone (red solid).
Note the systematic offset to lower \mbh\ at a fixed \sigmastar\ 
for the megamaser disk galaxies. Symbols as in Figure \ref{fig:bhcent}. 
\label{fig:msigma}}
\vskip 2mm

Calculating the net offset from our best-fit \mbh-\sigmastar\ relation
for elliptical galaxies, we find $\Delta M_{\rm BH} = -0.60 \pm 0.14$
dex for the 20 megamaser disks, while we find no mean offset 
$\Delta M_{\rm BH} = -0.15 \pm 0.15$ dex for the non-maser spirals
(Fig.\ \ref{fig:mhist}). The maser and non-maser spirals are
significantly different in $(M_{\rm BH}/\sigmastar)^5$; the
Anderson-Darling test returns a probability $P=0.015$ that they are
drawn from the same distribution (Fig. \ref{fig:mhist}), even if we
focus on just the 18 non-S0 maser disk galaxies or the maser and
non-maser pseudobulge samples ($P=0.01$).  Finally, we examine the two
samples non-parametrically in two dimensions using the Cramer
Von-Mises test, and find that the two samples are different at 90\%
significance.

With such small numbers we see only a suggestive difference between
the maser and non-maser samples.  On the other hand, we see a similar
trend in the \mbh-$M_{\rm bul}$ relations
\citep{laeskeretal2016}. Thus, we briefly consider the possible
ramifications should this difference prove real. There are three
possibilities. First, the masers could trace the true
\mbh\ distribution, while the non-maser \mbh\ distribution is biased
due to the inability to resolve the sphere of influence of
$<10^7$~\msun\ BHs. Existing upper limits do not constrain this
possibility, as they are at higher BH mass
\citep{gultekinetal2011}. In this scenario, spiral (or low-mass)
galaxies do have a broad range of \mbh, as may be expected in
merger-driven scenarios \citep[e.g.,][]{jahnkemaccio2011} or if
\mbh\ never gets large enough to exert feedback on the galaxy
\citep[e.g.,][]{zubovasking2012}.

Second, the non-maser distribution could be
the real one, with masers preferentially occurring in
lower-\mbh\ systems due to a correlation between \mbh\ and disk size
\citep[][]{neufeldetal1994,vandenboschetal2016}. The third
possibility, which we view as highly unlikely, is that we may be
preferentially catching the maser galaxies as they grow toward the
\mbh-\sigmastar\ relation. Given that BH growth episodes are likely
quite short compared to the growth times of bulges (many Gyr), we do
not favor this third scenario \citep{greeneetal2010}. We urgently need
more observations of both spiral and elliptical galaxies at low mass,
using multiple methods, to determine the full distribution of \mbh\ at
low galaxy mass.  

\vbox{ 
\vskip +1mm
\hskip +1mm
\includegraphics[width=0.45\textwidth]{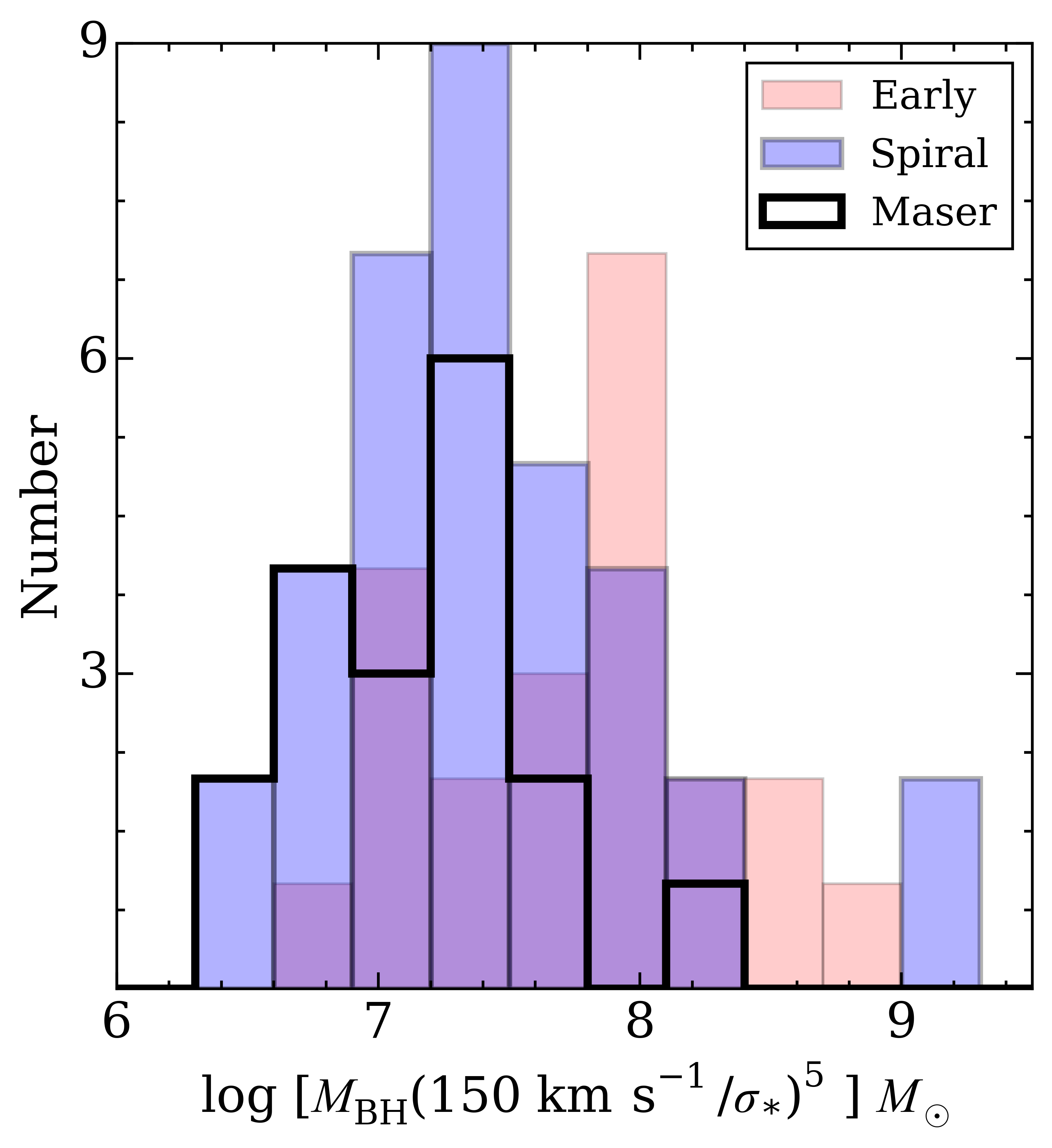}
}
\vskip -2mm
\figcaption[]{
The distribution of \mbh\ at fixed \sigmastar. 
Megamaser galaxies (thick black) 
are offset to lower \mbh\ than the full sample of spirals 
(blue filled) or the early-type galaxies (red filled).
\label{fig:mhist}}
\vskip 2mm

\section{Summary}

We present new stellar velocity dispersion measurements for seven
megamaser disk galaxies, and revisit galaxy-BH scaling relations that
incorporate these new measurements. In addition to \sigmastar, we
investigate total stellar mass $M_{\ast}$, and a non-parametric measure of
central stellar mass density within 1 kpc, $M_{\rm 1 kpc}$.  We have
also recently examined bulge mass \citep{laeskeretal2016}, and
halo mass \citep{sunetal2013}.

The \mbh\ in megamaser disk galaxies span a large range for 
any fixed galaxy property, while the maser and non-maser
dynamical measurements have different distributions of \mbh\ at a
fixed galaxy property.  To determine the real distribution of \mbh\ in
this low-mass regime we must push the sphere-of-influence limit
downwards for non-maser \mbh. 
We hope that with ALMA \citep[e.g.,][]{barthetal2016} 
and thirty-meter--class optical/NIR
telescopes \citep[e.g.,][]{doetal2014}, we will fill in low-mass
galaxies with non-maser dynamical measurements to address the
full distribution of \mbh\ in low-mass galaxies.

\acknowledgements

We thank the referee for a constructive report that significantly
improved this manuscript.  J.E.G. thanks A. Barth, K. Gultekin,
S. Tremaine, and R. van den Bosch for useful discussions.
J.E.G. acknowledges funding from NSF grant AST-1310405.  ACS
acknowledges support from NSF grant AST-1350389. This research has
made use of the NASA/IPAC Extragalactic Database (NED) which is
operated by the Jet Propulsion Laboratory, California Institute of
Technology, under contract with the National Aeronautics and Space
Administration.

\end{document}